%% file: 42669corr.tex
\documentclass{aa}

\usepackage{lmodern}
\usepackage[english]{babel}
\usepackage[T1]{fontenc}
\usepackage{graphicx}
\usepackage{caption}
\usepackage{amsmath, amssymb, amsfonts}
\usepackage{layout}
\usepackage{txfonts}
\usepackage{geometry}
\usepackage{subfigure}
\usepackage{wasysym}
\usepackage{multicol}
\usepackage{multirow}
\usepackage{chemfig}
\usepackage{arydshln}
\usepackage{mathtools}
\usepackage{natbib}
\bibpunct{(}{)}{;}{a}{}{,} 
\usepackage{xcolor,colortbl}
\usepackage{subfigure}
\usepackage{xspace}

\newcommand{\csixtyplus}{C$_{60}^{+}$\xspace}

\begin{document}

   \title{C$_{60}^+$ diffuse interstellar band correlations and \\ environmental variations}
   
   \titlerunning{$C_{60}^+$ DIBs correlations and environmental variations}
   \authorrunning{Schlarmann et al.}

   \author{Leander Schlarmann
          \inst{1,2,}\thanks{\email{leander.schlarmann@chello.at}}
          \and
          Bernard Foing \inst{2}
          \and
          Jan Cami \inst{3,4,5}
          \and
          Haoyu Fan \inst{3,4}
          }

   \institute{Department of Astrophysics, University of Vienna, Türkenschanzstrasse 17,  1180 Vienna, Austria. 
         \and
             ILEWG EMMESI EuroMoonMars Earth Space Innovation,  Leiden Observatory, Universiteit Leiden. Huygens Laboratory / J.H. Oort Building. Niels Bohrweg 2. NL-2333 CA Leiden. The Netherlands.
         \and
                 Department of Physics and Astronomy, The University of Western Ontario, London, ON N6A 3K7, Canada.
                 \and
                 Institute for Earth and Space Exploration, The University of Western Ontario, London, ON N6A 3K7, Canada.
                 \and
                 SETI Institute, 189 Bernardo Ave, Suite 100, Mountain View, CA 94043, USA. 
             }

   \date{Received November 15, 2021; accepted November 25, 2021}

 
  \abstract
   {The diffuse interstellar bands (DIBs) are absorption features seen in the spectra of astronomical objects that arise in the interstellar medium. Today, more than 500 DIBs have been observed, mostly in the optical and near-infrared wavelengths. The origin of the DIBs is unclear; only ionised buckminsterfullerene, \csixtyplus, has been identified as a viable candidate for two strong and three weaker DIBs. }
   {We investigate the correlations between the strengths of the two strongest C$_{60}^+$ DIBs as well as their environmental behaviour.}
   {We analysed measurements of the strengths of the two C$_{\text{60}}^{+}$ DIBs at 9577 and 9633~\AA\ for 26 lines of sight. We used two different methods, including Monte Carlo simulations, to study their correlations and the influence of measurement errors on the correlation coefficients. We examined how the strength of the C$_{\text{60}}^{+}$ DIBs changes as a result of different environmental conditions, as measured by the concentration of H/H$_2$ and the strength of the ambient UV radiation.}
   {In contrast to results recently reported by \cite{Galazutdinov2021}, we find a high correlation between the strengths of the C$_{\text{60}}^{+}$ DIBs. We also discovered that the behaviour of the correlated C$_{60}^+$ bands is quite distinct from other DIBs at 5780, 5797, and 6203~\AA\ in different environments. }
   {}


   \maketitle
%
\parindent 0pt

\section{Introduction}
\input{targettable}

In the last century, more than 500 diffuse interstellar bands (DIBs) have been found arising from the interstellar medium \citep[ISM;][]{2019ApJ...878..151F}. The first two DIBs were found by Mary Lea Heger in 1919, who observed two broad features around 5797 and 5780~\AA\ that originate from the ISM (\citealt{1922LicOB..10..146H}; see also \citealt{Herbig_1995} for a review).
The carrier molecules that give rise to these spectral features are unknown for most of the DIBs, with one exception: C$_{60}^+$. 

\cite{Foing_Ehrenfreund_1994} first discovered two relatively strong and broad interstellar features near 9577 and 9633~\AA. They proposed that these features were carried by ionised buckminsterfullerene, C$_{\text{60}}^{+}$, based on two strong absorption bands recorded for \csixtyplus embedded in rare gas matrices \citep[d'Hendecourt et al. in][]{Joblin1992,Fulara1993}. Recently, five near-infrared DIBs were attributed to C$_{\text{60}}^{+}$ by \cite{Campbell_2015}, who used ion trap technology to record laboratory spectra of C$_{\text{60}}^{+}$-He complexes and C$_{\text{60}}^{+}$ embedded in He droplets. These measurements confirmed the proposition that the 9633 and 9577 DIBs are due to C$_{\text{60}}^{+}$. Additionally, they reported three weaker bands at 9428, 9365, and 9348~Å that showed up in the laboratory data. 

Today, the identification of \csixtyplus is supported by an impressive array of observational and experimental studies \citep[see][for a review]{Linnartz_2020}. However, there has been some debate in the literature about this assignment due, at least partly, to several factors that complicate the comparison between the laboratory data and astronomical observations. First, all \csixtyplus features lie in spectral regions that are heavily contaminated by telluric water vapour lines, making their detection and proper characterisation difficult. This is especially hard for the weaker bands, but even the two stronger DIBs at 9633 and 9577~\AA\ suffer from telluric contamination, which results in uncertainties on the measurements \citep{Foing_Ehrenfreund_1994, Foing_1997, Galazutdinov2000, Galazutdinov2021}. \citet{Cordiner_2017, Cordiner_2019}, however, used the Hubble Space Telescope Imaging Spectrograph (HST/STIS) to observe several targets that exhibit the 9577 and 9632 DIBs. The use of space telescopes eliminates the need for error-prone telluric correction methods, at the cost of some spectral resolution. This allowed these authors to confirm the presence of the weaker C$_{\text{60}}^{+}$ DIBs in several targets; however, their observations did not include the wavelength range covering the 9632 \AA\ DIB. The second complicating factor is that for some spectral types the 9633 DIB overlaps with a stellar Mg~{\sc II} line, which can affect equivalent width (EW) measurements. When studying the \csixtyplus DIBs, it is thus necessary to study objects for which contamination is not an issue (either because the stellar line is not present or because the stellar radial velocity moves it well away from the DIB) or, alternatively, provide a good estimate for the strength of the stellar line, which is not a straightforward task. 

A key issue that has been used to challenge the \csixtyplus identification is based on correlation studies between the \csixtyplus DIBs. Correlation studies have a long history in DIB research, dating back to \citet{1973AJ.....78..913S}. Researchers have used correlation studies to investigate the nature of the DIB carriers by comparing the DIB EWs to different interstellar quantities, such as reddening E$_{B-V}$ and other extinction parameters \citep[e.g.][]{1973AJ.....78..913S, 2005MNRAS.358..563M, 2011ApJ...727...33F, 2012A&A...544A.136R, 2013A&A...555A..25P}, the column densities of interstellar atoms and small molecules \citep[e.g.][]{1992MNRAS.258..693K,2004MNRAS.355..169G,2011ApJ...727...33F}, and the strengths of other DIBs \citep[e.g.][]{1973AJ.....78..913S, Cami_1997, 1999A&A...351..680M, 2010ApJ...708.1628M, 2012PASJ...64...31X, Fan_2017, 2020MNRAS.496.2231B, 2021MNRAS.507.5236S}, often with the aim to group DIBs in families that could have similar (or even the same) carriers \citep{1987ApJ...312..860K, 1989A&A...218..216W, Cami_1997, 2003MNRAS.338..990W}. 

At the base of many of these correlation studies is the idea that if two DIBs arise from the same carrier, their EWs should in principle exhibit a perfect correlation since they both arise from the electronic ground state. Based on this assumption, \citet{Galazutdinov2021} state that the two strong bands at 9577~\AA\ and 9632~\AA\ should show a perfect correlation (to within uncertainties). According to their measurements and their analysis, however, the two bands in fact show a remarkably poor correlation, even accounting for measurement uncertainties, which led them to the conclusion that the two DIBs cannot be due to the same carrier and hence cannot be due to \csixtyplus. This is a surprising result in view of the good spectral matches with the laboratory experiments and begs a careful and critical examination. 

In this paper we compose a dataset from different literature sources to study the correlations of the two C$_{60}^+$ bands with each other as well as with several environmental parameters.

\section{Data}
The data we use in this paper include EW measurements of the 9577 and 9632~\AA\ DIBs and environmental parameters for 26 lines of sight that have been collected from different sources in the literature. We summarise them in Table \ref{data}. The environmental data, in particular the colour excess E$_{B-V}$, the column densities of neutral hydrogen and molecular hydrogen, N(H) and N(H$_2$), and the molecular fraction of hydrogen, f$_{H_2}$, were taken from \cite{Fan_2017}, who studied the behaviour of eight relatively strong DIBs in different interstellar environments. In the \cite{Fan_2017} dataset, most of the spectra were taken with the ARC echelle spectrograph (ARCES) on the 3.5 m telescope at the Apache Point Observatory (APO). \citet{Fan_2017}, however, did not include the 9577 and 9633~\AA\ DIBs. Therefore, most of the EWs for the 9633 and 9577 DIBs were taken from \cite{Galazutdinov2021}, who collected spectra of these bands using several high resolving power echelle spectrographs. Furthermore, for the line of sight toward HD~43384, the EWs were taken from \cite{Jenniskens_1997}, who used the 1.4m ESO Coudé Auxiliary Telescope (CAT) and the Coudé
Echelle Spectrometer (CES); for BD~+63~1964, we used the values from \cite{Cordiner_2017} obtained with HST/STIS. 

For HD~183143, the EWs for the C$_{60}^+$ bands were examined in different studies, with a rather large difference between the reported values. \cite{Foing_1997} report 266~m\AA\ and 311~m\AA\ for the EWs of the 9577 and 9633 DIBs, respectively, while \cite{Cox_2014} found 260~$\pm$~5~m\AA\ and 263~$\pm$~3~m\AA\ and \cite{Galazutdinov2021} list 105~$\pm$~20~m\AA\ and 300~$\pm$~20~m\AA. The disagreement in the values for the 9633 DIB could be (partly) due to the different methods used for the Mg II line correction. For the 9577~\AA\ DIB, any discrepancies are most likely due to telluric correction methods. The standard technique for the removal of telluric water vapour lines is the division of the reddened star spectrum by one from an un-reddened star of a similar spectral type observed through the same air mass. For the purposes of this paper, we chose to use the EWs determined by \cite{Cox_2014} for HD~183143.

\section{Methods}
For our correlation studies, we used the Pearson linear correlation coefficient, $r$. As one can see in Eq.\ (\ref{pears}), \begin{equation}
r = \frac{\smashoperator[r]{\sum_{i=1}^{n}}\left(x_i - \bar{x} \right)\left(y_i - \bar{y} \right)}{\sqrt{\smashoperator[r]{\sum_{i=1}^{n}}\left(x_i - \bar{x} \right)^2}\sqrt{\smashoperator[r]{\sum_{i=1}^{n}}\left(y_i - \bar{y} \right)^2}} \label{pears}
,\end{equation}
this coefficient is described by the ratio between the covariance of two variables, $x_i$ and $y_i$, and the product of their standard deviation. The result of this equation always gives a value between -1 and 1, where $r=$1 (or $r=-1$) indicates a strong positive (negative) relationship between the two variables and $r=$ 0 indicates no relationship at all.

Pair-wise comparisons between different quantities can help in identifying general relationships. To include the measurement error in the correlation calculations, we also calculated the weighted correlation coefficient $r_w(x,y;w)$ using the weighted covariance $\text{Cov}(x,y;w)$ at a given weight, $w$: 
\begin{equation}
\text{Cov}(x,y;w)=\frac{\sum_i w_i \, (x_i - \frac{\sum_i w_i x_i}{\sum_i w_i})(y_i - \frac{\sum_i w_i y_i}{\sum_i w_i})}{\sum_i w_i}
.\end{equation}

The weighted correlation coefficient $r_w(x,y;w)$ can then be calculated with Eq.\ (\ref{w_pears}):\begin{equation}
r_w(x,y;w)=\frac{\text{Cov}(x,y;w)}{\sqrt{\text{Cov}(x,x;w)\text{Cov}(y,y;w)}}\label{w_pears}
.\end{equation} 

 For the purpose of this study, the measurement error $\sigma$ was taken as a weight in the form of $1/\sigma^2$.
We also adopted an independent approach to assess the influence of measurement errors by performing Monte Carlo simulations in which we first constructed a few thousand new datasets. Each of these datasets was constructed from the original measurements, but we added a random value to each of the measurements drawn from a normal distribution with a standard deviation equal to the measurement error. These thousand datasets thus represent a thousand realisations of our original measurements. For each dataset, we calculated the Pearson correlation coefficient, $r$. We then determined the mean and the standard deviation of the correlation coefficients corresponding to each of the thousands of datasets. 

\section{Results and discussion}
\begin{figure}[t]
    \centering
    \includegraphics[width=0.49\textwidth]{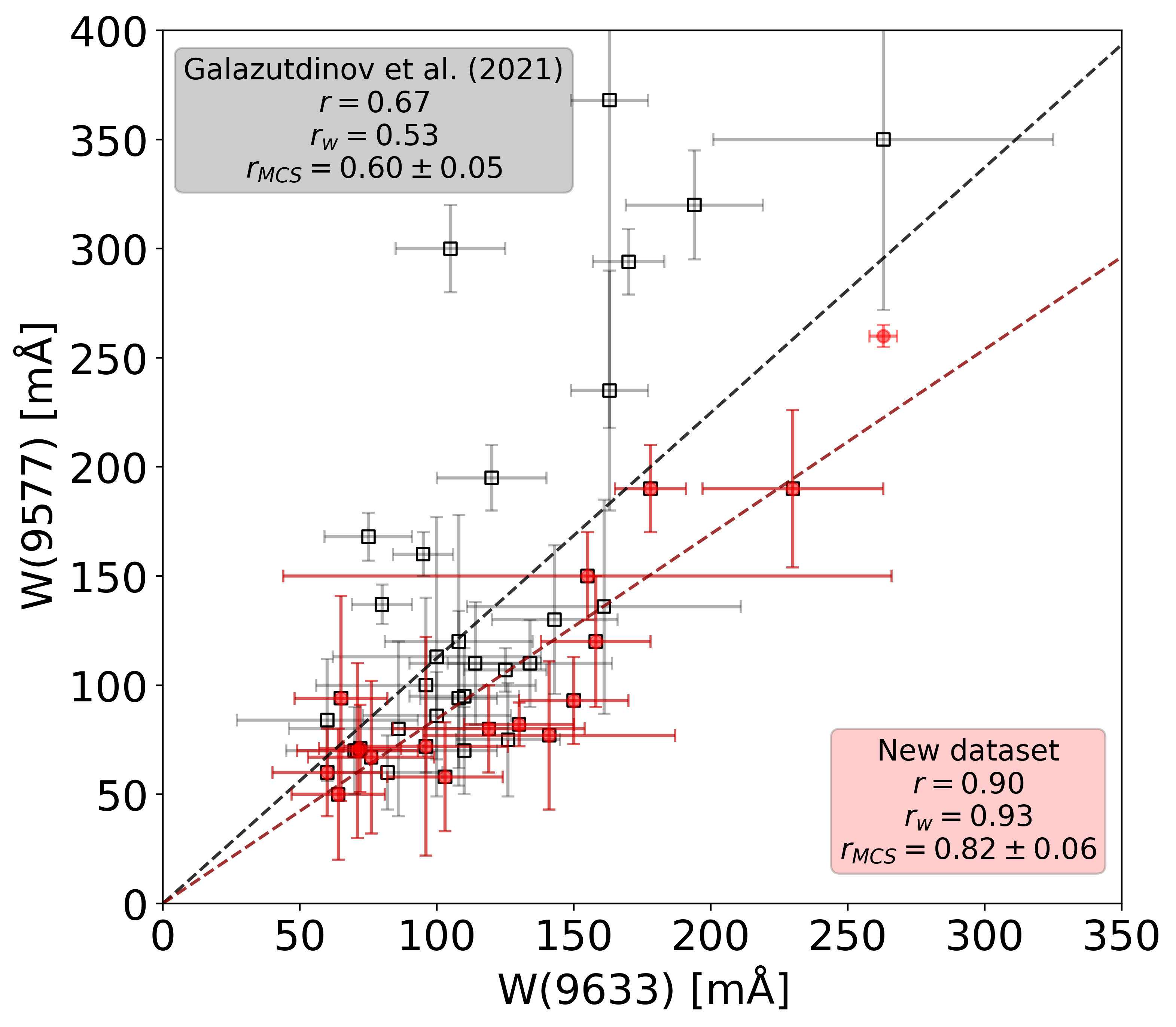}
    \caption{Equivalent width measurements of the 9577 DIB versus those of the 9633 DIB and the various resulting correlation coefficients. The red data points represent the measurements from Table~\ref{data}. The open black squares show the EWs of 43 lines of sight from \cite{Galazutdinov2021}, with measurements for both \csixtyplus DIBs. We note the relatively high correlation coefficients for our dataset.\label{Fig1_1}}
\end{figure}

\subsection{Correlation between the 9577 and 9633\AA~DIBs }
We first investigate the correlation between the two strong \csixtyplus DIBs. It should be noted that in our sample of stars we have measurements for the 9577 DIB for 24 objects and for the 9633 DIB for 20 objects. 

As one can see in Fig. \ref{Fig1_1}, we find a relatively high correlation coefficient of $r=0.90$ for these two DIBs. For this dataset, the weighted correlation coefficient, $r_w=0.93$, is even slightly larger than the regular Pearson correlation coefficient, $r$. Using Monte Carlo simulations, where $r$ was calculated 4000 times, the calculated correlation coefficient is slightly smaller but still fairly high, at r$_{MCS}=0.82\pm0.06$. 

This result is inconsistent with \cite{Galazutdinov2021}, who report a poor mutual correlation of only $r=0.37$ between the EWs of the strong C$_{60}^+$ DIBs. However, it should be noted that in their study they used observations of 62 lines of sight obtained using several high resolving power echelle spectrographs, with measurements of the EWs for both \csixtyplus DIBs for 43 lines of sight. They state that they calculated the correlation coefficients with the measurement error $\sigma$ taken as a weight in the form of $1/\sigma^2$. Because of the different results of the correlation calculations of this dataset and that of \cite{Galazutdinov2021}, we recalculated the correlation coefficient with their published data (the black data points in Fig.~\ref{Fig1_1}) using different methods. As one can see in Fig. \ref{Fig1_1}, we find instead a correlation coefficient of $r=0.67$, a weighted correlation coefficient $r_w=0.53$ (where we used 1/$\sigma^2$ as the weight), and a Monte Carlo result of $r_{MCS}=0.60\pm0.05$. While these values are smaller than ours, none of them come close to their reported value of $r=0.37$. 

It is important in this context to realise that if two variables are intrinsically highly correlated, measurement uncertainties will always reduce the correlation between them. Thus, in the presence of noise, even auto-correlations can yield values significantly lower than unity. To get a sense of the magnitude of this effect on our calculations, we calculated the `self-correlation' of both the 9633 and 9577 DIBs for our dataset as well as that from \cite{Galazutdinov2021}. Not surprisingly, the correlation coefficient, $r$, and the weighted correlation coefficient, $r_w$, have a value of 1 for both datasets. However, when performing such a correlation with a Monte Carlo simulation, the achieved correlation is 0.87-0.93 -- of the same magnitude as the values for the mutual correlations. This shows that even if two DIBs were perfectly correlated intrinsically, they would yield an effective correlation of about 0.9 due to measurement errors. \cite{Cami_1997} found that this systematic `decorrelation bias' of 1.8 $\sigma$ in the auto-correlation has to be considered to check the reliability of the calculated correlation. Given the large uncertainties on the EW measurements, the correlation we find between the two \csixtyplus DIBs is thus an indication of an even tighter intrinsic correlation, and possibly even a perfect correlation. 

It is important in this context to point out that the idea that absorption bands originating from the same electronic ground state should show a perfect correlation has indeed long been a key assumption that has driven many correlation studies. However, this assumption does not always hold, and \csixtyplus is one such exception. Indeed, \citet{Lykhin:C60+} carried out detailed computations to elucidate the electronic structure of \csixtyplus relevant to the DIBs. They conclude that the bands at 9577 and 9632~\AA\ arise from excitation in the non-Franck-Condon region of the ground electronic state. They further suggest that ro-vibrational coupling can affect the Franck-Condon factors and therefore the intensities of the bands. If true, this would then imply that the ratio of the two DIB strengths could be used as a measure of the interstellar temperature. In any event, it illustrates that the band ratio does not need to be constant, even for transitions from the same species and the same electronic ground state. 

\subsection{Behaviour of C$_{60}^+$ DIBs with H and H$_2$ column densities}
\begin{figure}[t]
    \centering
    \includegraphics[width=0.49\textwidth]{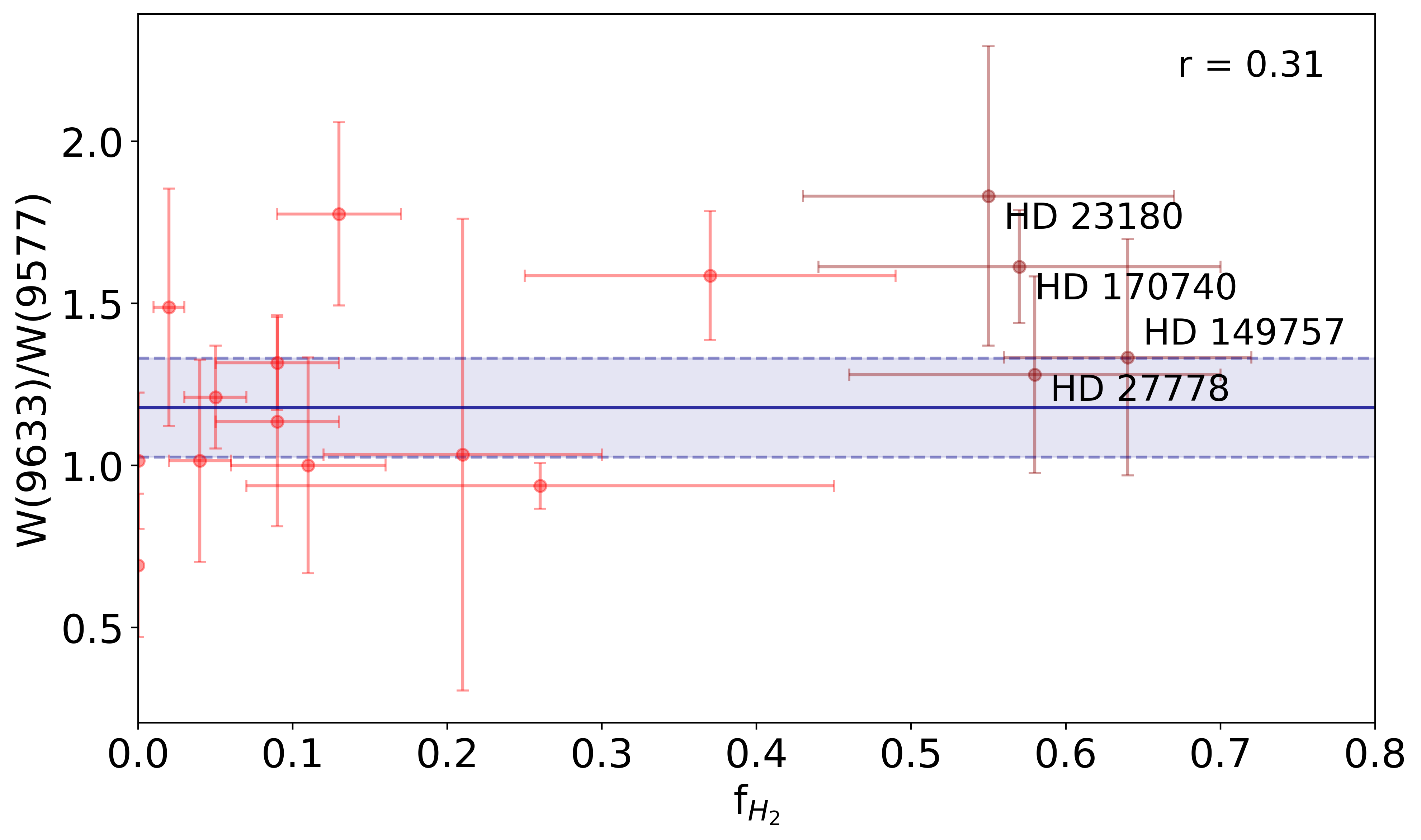}
    \caption{W(9633)/W(9577) ratio as a function of f$_{H_2}$.}
    \label{Fig1_2}
\end{figure}

The column densities of H and H$_2$ were measured in a number of our targets, as reported by \cite{Fan_2017}. For a homogeneous line of sight, f$_{\rm H}$ and f$_{\rm H_2}$ represent the average atomic and molecular fraction. 
However, interstellar lines of sight are heterogeneous, including the conditions that represent the outside, inside, and edge of interstellar clouds (and sometimes of multiple clouds). Ionised buckminsterfullerene is mostly related to conditions in the outer edge of clouds, where there are sufficient photons with energy $\ge$~7~eV to ionise C$_{60}$, in balance with the local electron density, $n_e$. We note that in H II regions with an abundance of photons $\ge$~13.6 eV, C$_{60}^+$ would be further ionised to C$_{60}^{++}$ (the ionisation potential being 11.35~eV) but nonetheless would not break its cage structure. 

The molecular hydrogen fraction, f$_{H_2}$, can be interpreted as an indicator of the typical local hydrogen density in the main interstellar clouds in each line of sight \citep{Fan_2017}. As H$_2$ dissociates under the influence of UV radiation, f$_{H_2}$ can also be used as an approximation of the strength of the local UV field \citep{Cami_1997}. 
In Fig.~\ref{Fig1_2}, the ratio W(9633)/W(9577) is indicated versus this average f$_{H_2}$ indicator, showing values in the range 1.0-1.5 at f$_{H_2}$<0.3 and 1.3-1.8 for f$_{H_2}$>0.35. 
At first approximation, there is no significant variation in the ratio across our sightlines, taking into account the noise in data and mixing of heterogeneous interstellar conditions. 

\begin{figure*} [t]
    \subfigure[Normalised EWs of the C$_{60}^{+}$ DIBs as functions of f$_{H}$\label{Fig1_22}]{\includegraphics[width=0.49\textwidth]{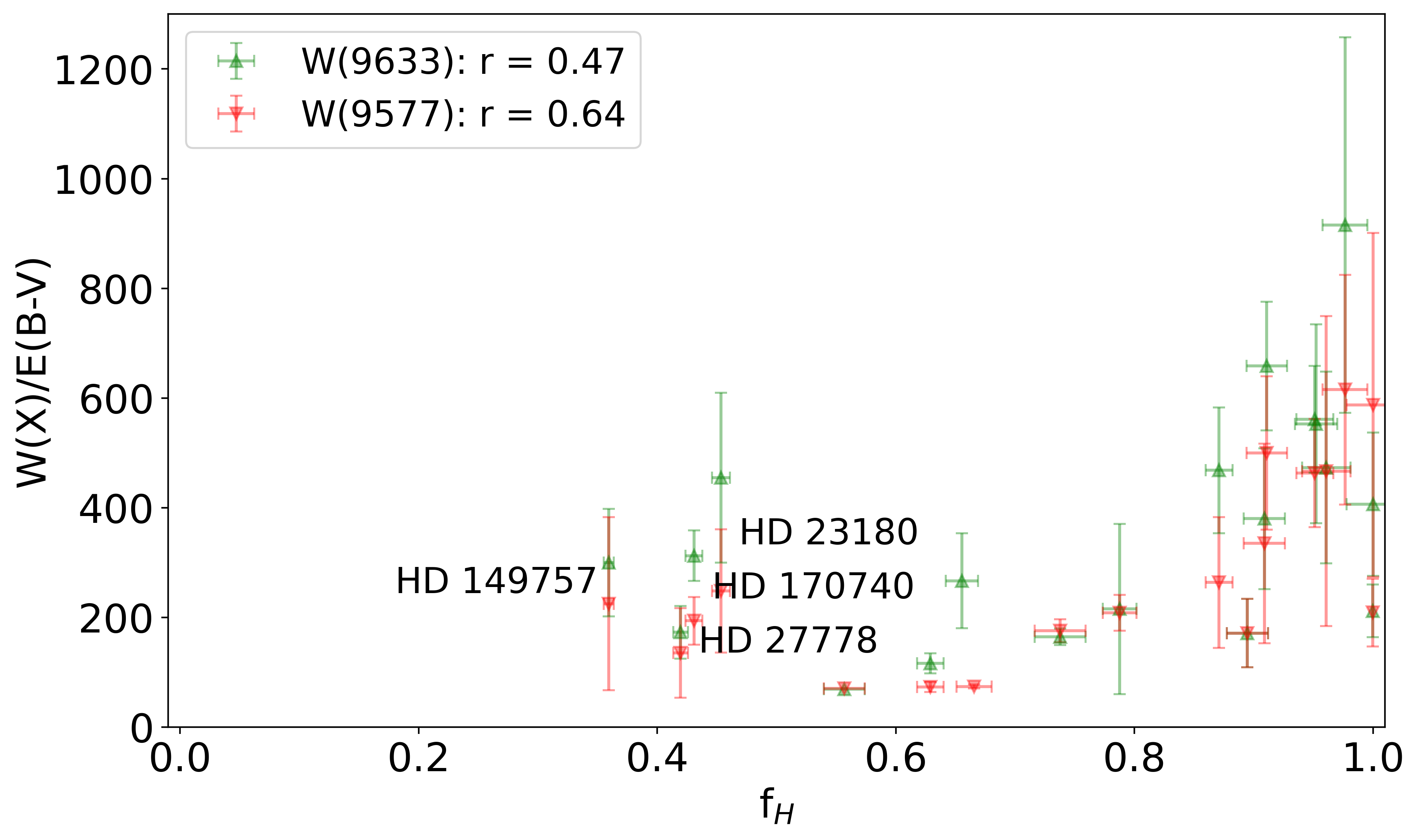}}
    \subfigure[Normalised EWs of the other DIBs as functions of f$_{H}$\label{Fig1_23}]{\includegraphics[width=0.49\textwidth]{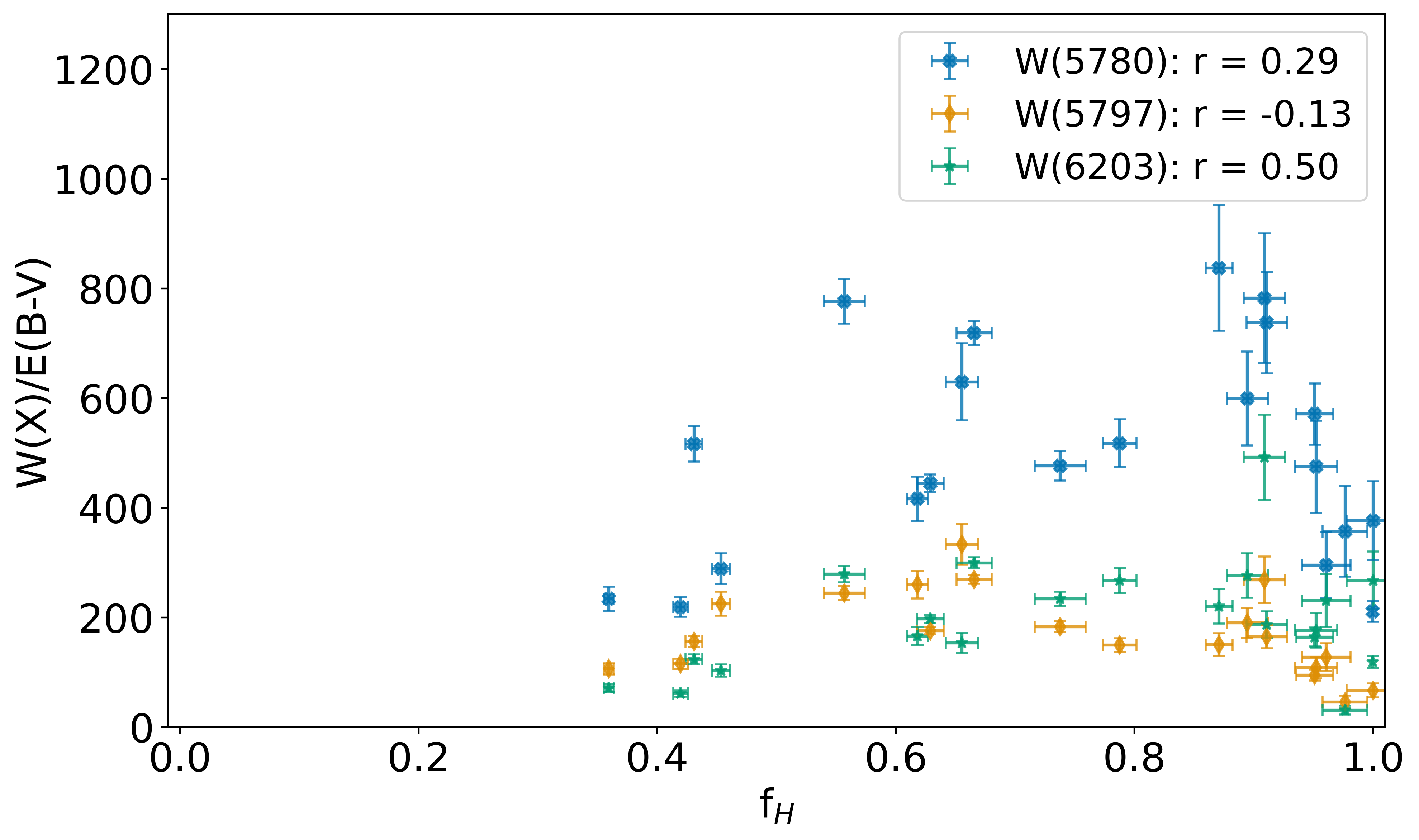}}
        \subfigure[Normalised EWs of the C$_{60}^{+}$ DIBs as functions of E$_{B-V}$\label{Corr_EBV_C60}]{\includegraphics[width=0.485\textwidth]{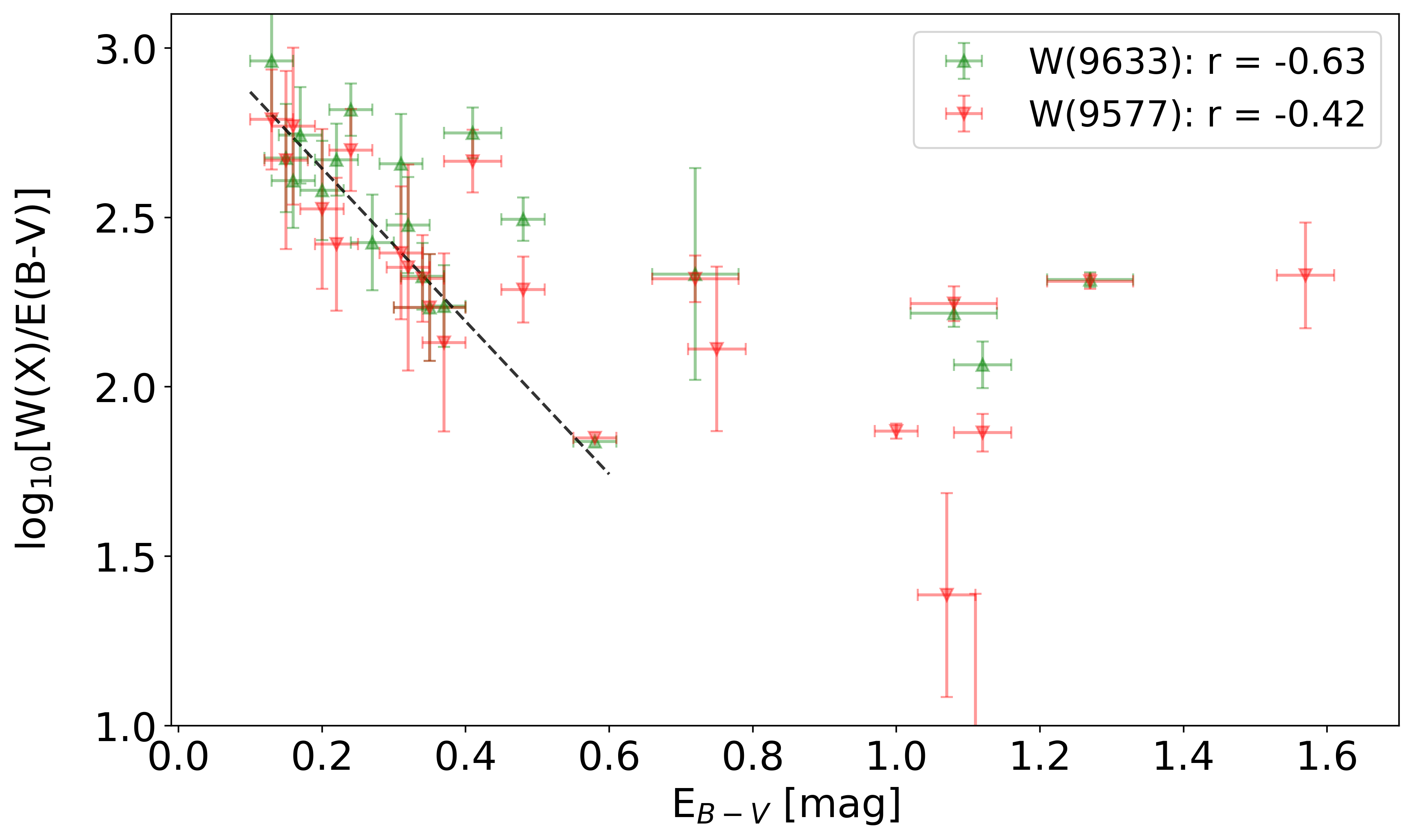}}
        \hspace{0.8em}
    \subfigure[Normalised EWs of the other DIBs as functions of E$_{B-V}$\label{Corr_EBV_other}]{\includegraphics[width=0.485\textwidth]{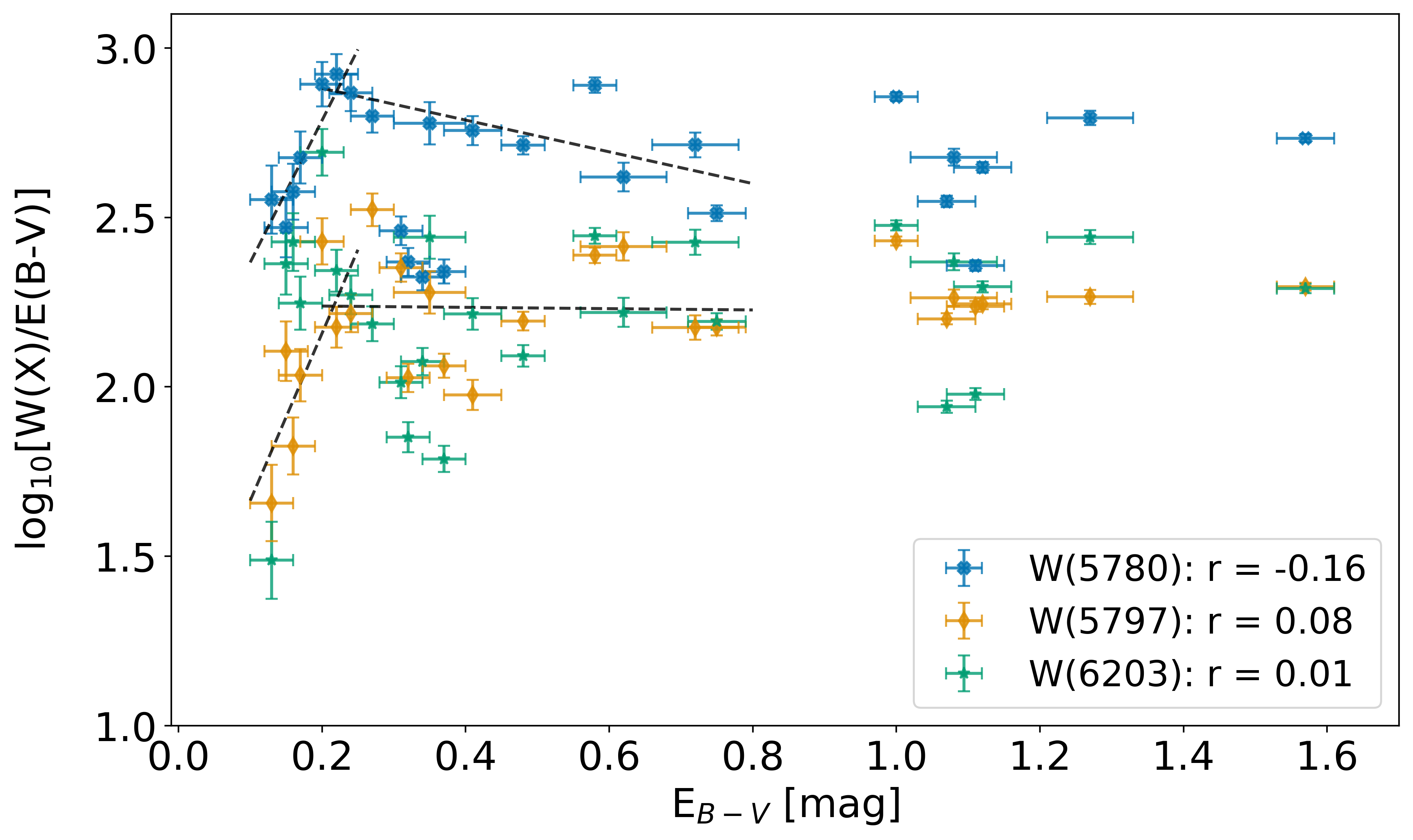}}
    \caption{Correlation studies for the C$_{60}^+$ bands. The normalised EWs of the two C$_{\text{60}}^{+}$ DIBs and the DIBs at 6203, 5780, and 5797 \AA\ are plotted against the fraction of neutral hydrogen f$_H$ (top) and the colour excess E$_{B-V}$ (bottom). We note the very specific behaviour of C$_{\text{60}}^{+}$ compared to other DIB carriers.}
    \label{Corr_Norm_W}
\end{figure*}

In Fig. \ref{Fig1_22} the normalised EWs of the 9633 and 9577 DIBs are plotted against the fraction of neutral hydrogen, f$_{H}$. Using Eq.~\ref{EqfH2} the neutral hydrogen fraction, f$_{H}$, can be calculated from the column densities of neutral hydrogen and molecular hydrogen, N(H) and N(H$_2$):
\begin{equation}
\text{f}_{H} = \frac{\text{N(H)}}{\text{N(H)}+2\,\text{N(H}_2\text{)}}\label{EqfH2}
.\end{equation}
The normalised EWs of the C$_{60}^+$ DIBs experience a `u-shaped' behaviour, seen in Fig. \ref{Fig1_22}. While W(DIB)/E(B-V) increases strongly with f$_{H}$ at high f$_{H}$, the lines of sight HD~23180, HD~170740, HD~149757, and HD~27778 are relatively strong at around f$_{H}=0.4$. This would not be expected for ionised molecules such as C$_{60}^+$. The reason could be the presence of multiple velocity components in the line of sight, indicating the presence of clouds with different compositions. Therefore, the molecular fraction of H$_2$ could arise from a different region where C$_{60}^+$ is not that abundant. For example, HD 170740 has two evident components in the profiles of He~I and C~II \citep{Galazutdinov_2017}. Furthermore, a stellar Mg II~line could influence the accuracy of some of the measured EWs of the 9633 DIB. \cite{2017ApJ...843...56W} found close spectral standards to be superior to corrections based on line profiles from model atmosphere calculations introduced by \cite{Galazutdinov_2017} in correcting for contamination by this stellar Mg II line.

We expect C$_{60}^+$ to be present in the outer edges of clouds, where more UV-processed material can be found. 
In Table \ref{data2} the normalised EWs of the DIBs at 6203~\AA, 5780~\AA,\ and 5797~\AA\ from \cite{Fan_2017} are shown for the 26 investigated lines of sight. These DIBs have been subject to many past studies, and the measurement errors on their EWs are very small. In Fig. \ref{Fig1_23} the normalised EWs of these DIBs are plotted against f$_{H}$ once again. In contrast to the C$_{60}^+$ DIBs, here the DIB strengths increase with f$_{H}$ until a maximum is reached, after which the DIB strengths decrease with f$_{H}$. This rise and fall of the normalised DIB strength in Fig.~\ref{Fig1_23} is quite different from the physical behaviour of the correlated C$_{60}^+$ bands in Fig.~\ref{Fig1_22}.

\subsection{Correlations with E$_{B-V}$ and hydrogen fraction}

The colour excess E$_{B-V}$ is calculated from the difference between the observed colour index of an object (B-V)$_{\text{observed}}$ and the intrinsic colour index (B-V)$_{\text{intrinsic}}$ predicted from its spectral type. Thus, the optical depth of interstellar dust can be related to E$_{B-V}$.
In Fig. \ref{Corr_EBV_C60} the relation of the two normalised C$_{60}^{+}$ DIBs to E$_{B-V}$ is shown. The dataset of Table \ref{data} contains two outliers, Cyg OB2 12 and HD 168607, where E$_{B-V}$ is larger than 1.5. To study the behaviour and ionisation properties of the DIBs at low E$_{B-V}$, we focused on the lines of sight with E$_{B-V}<1.6$. We note that the ionisation potential (IP) of C$_{60}^{+}$ is 11.35 eV \citep{PhysRevA.93.033401}.
\cite{Sonnentrucker_1997} developed a simple photoionisation model for the DIB carriers based on the DIB variations at low reddening. Their model enables us to estimate the DIB IPs from the slope of the variation in the normalised DIB strength as a function of reddening.
The prediction for C$_{60}^{+}$ would be a slope of 3.1 for log(W/E$_{B-V}$) versus E$_{B-V}$. For ionised carriers, such as C$_{60}^{+}$, they are predicted to become neutral in higher density regions. Thus, the normalised EWs of the corresponding DIBs would decrease with E$_{B-V}$. This coincides with our results in Fig. \ref{Corr_EBV_C60}, where the negative slope for E$_{B-V}$ values between 0.1 and 0.6 indicates that more C$_{60}^{+}$ is transformed into C$_{60}$. The regime where  C$_{60}^{+}$ might be ionised to the doubly charged fullerene cation C$_{60}^{++}$ cannot be seen in Fig. \ref{Corr_EBV_C60}, and yet this might happen at E$_{B-V}< 0.1$ for single clouds.  This result suggests that ionised buckminsterfullerenes reside mostly outside the clouds. 

In Fig. \ref{Corr_EBV_other} the normalised EWs of the 5780, 5797, and 6203 DIBs are plotted against E$_{B-V}$. Here a rise and fall can be seen for low values of E$_{B-V}$. This lambda-shaped behaviour was also reported by \cite{Fan_2017} for the 5780 and 5797 DIBs. We note that the negative slope of -1 for log(W5780/E) for E$_{B-V}$ = 0.3 to 0.7 can be explained by a simple ionisation model where the DIB carrier column density is carried in the edge of a cloud \citep{Sonnentrucker_1997}. 
However, this slope might be affected if other factors such as hydrogenation (vs. H/UV) or electron density balance are affected within the clouds \citep{Vuong2000}.
For C$_{60}^{+}$ the  negative slope of -2 for $\log\left(W/\text{E}_{B-V}\right)$ versus E$_{B-V}$ for $0.2 < \text{E}_{B-V} <0.4$ is sharper than the -1 slope expected for a pure electron recombination effect (and as observed for $W(5780)$). This could indicate that C$_{60}^{+}$ is reduced in the inner edge of clouds due not only to the formation of C$_{60}$ but also possibly to other compounds, for instance via hydrogenation.

\begin{table*}[t]
\centering
\caption{\label{data2}Equivalent widths of the DIBs at 6203, 5780, and 5797 \AA\ for the 26 lines of sight from \cite{Fan_2017}.}
\resizebox{\textwidth}{!}{%
\begin{tabular}{cccccccc}
\hline
\hline
\multicolumn{8}{c}{} \\[-1em]
\rowcolor{white}
Name      & \multicolumn{1}{c}{W(6203) [m\AA]}                     & W(5780) [m\AA]              & W(5797) [m\AA]           & Name        & W(6203) [m\AA]                         & W(5780) [m\AA]                & W(5797) [m\AA]  \\
\hline
\multicolumn{8}{c}{} \\[-1em]
HD 22951  & 41.42   $\pm$  1.68    & 169.9  $\pm$  1.9 & 89.9  $\pm$ 1.1 & HD 147889   & 93.39  $\pm$ 1.70    & 377.4  $\pm$  3.3 & 169.9  $\pm$  0.7 \\
HD 23180  & 31.97   $\pm$  1.57    & 89.5   $\pm$  1.4 & 69.7  $\pm$ 0.4 & HD 148605   & 4.00   $\pm$ 0.50    & 46.4   $\pm$  1.0 & 5.9    $\pm$  0.7 \\
HD 27778  & 22.66   $\pm$  0.86    & 81.0   $\pm$  1.1 & 42.7  $\pm$ 0.7 & HD 149757   & 22.73  $\pm$ 0.92    & 74.8   $\pm$  1.4 & 34.0   $\pm$  0.7 \\
HD 36861  & 34.60   $\pm$  2.10    & 44.3   $\pm$  1.6 & 19.1  $\pm$ 0.6 & HD 167971   & 252.48 $\pm$ 2.86    & 514.3  $\pm$  5.9 & 197.7  $\pm$  1.3 \\
HD 37022  & 40.37   $\pm$  1.23    & 71.7   $\pm$  1.4 &                         & HD 168607   & 306.31 $\pm$ 5.94    & 850.3  $\pm$  9.0 & 310.1  $\pm$  1.6 \\
HD 40111  & 98.40   $\pm$  4.80    & 156.4  $\pm$  2.8 & 53.7  $\pm$ 2.5 & HD 169454   & 220.95 $\pm$ 2.64    & 497.7  $\pm$  2.0 & 196.7  $\pm$  1.6 \\
HD 43384  & 161.78  $\pm$  2.36    & 450.2  $\pm$  3.3 & 141.8 $\pm$ 1.1 & HD 170740   & 59.22  $\pm$ 2.29    & 247.8  $\pm$  1.5 & 75.0   $\pm$  0.2 \\
HD 54662  & 96.73   $\pm$  3.31    & 209.7  $\pm$  2.1 & 66.5  $\pm$ 0.5 & HD 183143   & 351.05 $\pm$ 1.31    & 789.5  $\pm$  8.4 & 234.0  $\pm$  0.6 \\
HD 57061  & 42.80   $\pm$  2.50    & 60.2   $\pm$  2.0 & 10.7  $\pm$ 0.5 & HD 190603   & 192.31 $\pm$ 3.53    & 372.7  $\pm$  2.7 & 107.7  $\pm$  0.9 \\
HD 143275 & 30.00   $\pm$  1.00    & 80.7   $\pm$  0.8 & 18.4  $\pm$ 0.5 & Cyg OB2 12  & 399.81 $\pm$ 4.74    & 843.0  $\pm$  8.5 & 355.9  $\pm$  3.1 \\
HD 144470 & 48.44   $\pm$  2.08    & 184.2  $\pm$  1.8 & 33.0  $\pm$ 1.0 & HD 204827   & 105.66 $\pm$ 1.68    & 253.2  $\pm$  1.6 & 191.7  $\pm$  0.5 \\
HD 145502 & 44.77   $\pm$  1.85    & 177.0  $\pm$  2.1 & 39.5  $\pm$ 1.0 & HD 208501   & 116.94 $\pm$ 1.54    & 244.1  $\pm$  2.3 & 112.4  $\pm$  1.3 \\
HD 147165 & 67.28   $\pm$  2.82    & 234.0  $\pm$  3.0 & 38.8  $\pm$ 1.3 & BD +63 1964 & 299.47 $\pm$ 4.59    & 718.4  $\pm$  4.6 & 269.4  $\pm$  0.9 \\
\hline
\hline
\end{tabular}
}
\end{table*}

\section{Conclusions}

We have analysed the C$_{60}^+$ DIBs at 9633 and 9577 \AA, performing correlation studies with different methods. We find a high correlation of $r=0.90$ between these two DIBs for a dataset of 26 lines of sight, in contrast to results by \cite{Galazutdinov2021} who found a poor mutual correlation of the C$_{60}^+$ DIBs and questioned the common origin of these DIBs. We have not been able to reproduce their results even when using their listed measurement values. A recent study by \cite{2021arXiv211105769N} also supports a close 9577 and 9633 \AA\ correlation.

Correlation studies of the C$_{60}^+$ DIBs with different environmental parameters show that the ratio of the two C$_{60}^+$ DIBs does not change significantly within errors for a very large range of conditions. 
A distribution of C$_{60}^+$ band ratios of 1-1.4 and 1.3-1.8 for an extreme range of conditions may be investigated with more accurate measurements and better modelling of intrinsic environment conditions. This could lead to a better understanding of possible changes in the band ratio and ratios that are slightly different from those measured in the laboratory. 

In lines of sight dominated by atomic H, we find a correlation between the normalised EWs of the C$_{60}^+$ DIBs and f$_H$. We find also a negative correlation of relative band strength at E$_{B-V}$ from 0.15-0.5, indicating the decrease in C$_{60}^+$ entering the edge of clouds due to electron and H recombination. We find that 
C$_{60}^+$  survives outside and in the outer edges of clouds, as opposed to many other DIB carriers that are depleted for single clouds thinner than E$_{B-V}$=0.2. Especially in the near-infrared wavelength range of the C$_{60}^+$ bands, it can be challenging to correct for telluric and stellar spectral lines in the lines of sight \citep{EDIBLES1}. We look forward to extending the analysis with ESO Diffuse Interstellar Bands Large Exploration Survey (EDIBLES) spectra \citep{Lallement_2018} in future studies.

\section*{Acknowledgements}
LS would like to thank the organisers of the Leiden/ESA Astrophysics Program for Summer Students (LEAPS) 2021. JC and HF acknowledge support from an NSERC Discovery Grant and a SERB Accelerator Award from Western University.

\addcontentsline{toc}{section}{Bibliography}
\bibliographystyle{aa}
\bibliography{Literatur}

\end{document}

%% file: targettable.tex
\begin{table*}[t]
\centering
\caption{\label{data}Data of the 9632 and 9577 Å DIBs and environmental data for 26 lines of sight.}
\resizebox{\textwidth}{!}{%
\begin{tabular}{cccccccc}
\hline
\hline
\multicolumn{8}{c}{} \\[-1em]
\rowcolor{white}
Name        & SpT       & E$_{\text{B-V}}$ [mag]  & $\log_{10}$N(H)  & $\log_{10}$N(H$_2$)  & f$_{H_2}$   &  \multicolumn{1}{c}{W(9633) [m\AA]} & W(9577) [m\AA] \\
\hline
HD 22951    & B0.5V     & 0.27   $\pm$ 0.03$^a$      & 21.04 $\pm$ 0.13$^a$    & 20.46 $\pm$ 0.14$^a$     & 0.34    $\pm$ 0.14$^a$   & 72   $\pm$ 22$^b$      &                   \\
HD 23180    & B1III     & 0.31   $\pm$ 0.03$^a$      & 20.82 $\pm$ 0.09$^a$    & 20.60 $\pm$ 0.12$^a$     & 0.55    $\pm$ 0.12$^a$   & 141  $\pm$ 46$^b$      & 77  $\pm$ 34$^b$      \\
HD 27778    & B3V       & 0.37   $\pm$ 0.03$^a$      & 20.95 $\pm$ 0.15$^a$    & 20.79 $\pm$ 0.06$^a$     & 0.58    $\pm$ 0.12$^a$   & 64   $\pm$ 17$^b$      & 50  $\pm$ 30$^b$      \\
HD 36861    & O8e       & 0.15   $\pm$ 0.03$^a$      & 20.81 $\pm$ 0.12$^a$    & 19.12 $\pm$ 0.14$^a$     & 0.04    $\pm$ 0.02$^a$   & 71   $\pm$ 22$^b$      & 70  $\pm$ 40$^b$      \\
HD 37022    & O6        & 0.34   $\pm$ 0.03$^a$      & 21.54 $\pm$ 0.11$^a$    & 17.55$^a$                & 0.00021$^a$         	  & 72   $\pm$ 15$^b$      & 71  $\pm$ 20$^b$      \\
HD 40111    & B0.5II    & 0.20   $\pm$ 0.03$^a$      & 21.03 $\pm$ 0.09$^a$    & 19.73 $\pm$ 0.14$^a$     & 0.09    $\pm$ 0.04$^a$   & 76   $\pm$ 23$^b$      & 67  $\pm$ 35$^b$      \\
HD 43384    & B3Ib      & 0.58   $\pm$ 0.03$^a$      & 21.27 $\pm$ 0.30$^a$    & 20.87 $\pm$ 0.14$^a$     & 0.44    $\pm$ 0.25$^a$   & 40 $\pm$ 18$^d$  & 41 $\pm$ 18$^d$           \\
HD 54662    & O7III     & 0.35   $\pm$ 0.05$^a$      & 21.23 $\pm$ 0.10$^a$    & 20.00 $\pm$ 0.14$^a$     & 0.11    $\pm$ 0.05$^a$   & 60   $\pm$ 20$^b$      & 60  $\pm$ 20$^b$      \\
HD 57061    & O9III     & 0.16   $\pm$ 0.03$^a$      & 20.80 $\pm$ 0.08$^a$    & 15.48 $\pm$ 0.14$^a$     & 0       $\pm$ 0$^a$      & 65   $\pm$ 17$^b$      & 94  $\pm$ 47$^b$      \\
HD 143275   & B0.3IV    & 0.17   $\pm$ 0.03$^a$      & 21.01 $\pm$ 0.08$^a$    & 19.41 $\pm$ 0.14$^a$     & 0.05    $\pm$ 0.02$^a$   & 94   $\pm$ 26$^b$      &                   \\
HD 144470   & B1V       & 0.22   $\pm$ 0.03$^a$      & 21.18 $\pm$ 0.09$^a$    & 20.05 $\pm$ 0.08$^a$     & 0.13    $\pm$ 0.04$^a$   & 103  $\pm$ 21$^b$      & 58  $\pm$ 25$^b$      \\
HD 145502   & B3V       & 0.24   $\pm$ 0.03$^a$      & 21.20 $\pm$ 0.12$^a$    & 19.89 $\pm$ 0.12$^a$     & 0.09    $\pm$ 0.04$^a$   & 158  $\pm$ 20$^b$      & 120 $\pm$ 30$^b$      \\
HD 147165   & B2III+O9V & 0.41   $\pm$ 0.04$^a$      & 21.38 $\pm$ 0.08$^a$    & 19.79 $\pm$ 0.12$^a$     & 0.05    $\pm$ 0.02$^a$   & 230  $\pm$ 33$^b$      & 190 $\pm$ 36$^b$      \\
HD 147889   & B2V       & 1.07   $\pm$ 0.04$^a$      & 21.80 $\pm$ 0.05$^a$    &                          & 0.48    $\pm$ 0.19$^a$   &    		& 26  $\pm$ 18$^b$      \\
HD 148605   & B2V       & 0.13   $\pm$ 0.03$^a$      & 20.66 $\pm$ 0.08$^a$    & 18.74 $\pm$ 0.14$^a$     & 0.02    $\pm$ 0.01$^a$   & 119  $\pm$ 35$^b$      & 80  $\pm$ 20$^b$      \\
HD 149757   & O9.5Ve    & 0.32   $\pm$ 0.03$^a$      & 20.69 $\pm$ 0.10$^a$    & 20.64 $\pm$ 0.06$^a$     & 0.64    $\pm$ 0.08$^a$   & 96   $\pm$ 30$^b$      & 72  $\pm$ 50$^b$      \\
HD 167971   & O8e       & 1.08   $\pm$ 0.06$^a$      & 21.60 $\pm$ 0.30$^a$    & 20.85 $\pm$ 0.12$^a$     & 0.26    $\pm$ 0.19$^a$   & 178  $\pm$ 13$^b$      & 190 $\pm$ 20$^b$      \\
HD 168607   & B9Ia      & 1.57   $\pm$ 0.04$^a$      &               	   			&                          &                          &                    & 335 $\pm$ 120$^b$     \\
HD 169454   & B1.5Ia    & 1.12   $\pm$ 0.04$^a$      & 21.51 $\pm$ 0.10$^a$    & 20.98 $\pm$ 0.13$^a$     & 0.37    $\pm$ 0.12$^a$   & 130  $\pm$ 20$^b$      & 82  $\pm$ 10$^b$      \\
HD 170740   & B2V       & 0.48   $\pm$ 0.03$^a$      & 21.04 $\pm$ 0.15$^a$    & 20.86 $\pm$ 0.08$^a$     & 0.57    $\pm$ 0.13$^a$   & 150  $\pm$ 20$^b$      & 93  $\pm$ 20$^b$      \\
HD 183143   & B7Iae     & 1.27   $\pm$ 0.06$^a$      &                    			&                          &                          & 263 $\pm$ 5$^c$		 & 260 $\pm$ 5$^c$     	 \\
HD 190603   & B1.5iAE   & 0.72   $\pm$ 0.06$^a$      & 21.41 $\pm$ 0.10$^a$    & 20.54 $\pm$ 0.13$^a$     & 0.21    $\pm$ 0.09$^a$   & 155  $\pm$ 111$^b$     & 150 $\pm$ 20$^b$      \\
Cyg OB2 12  & B5Ie      & 3.31   $\pm$ 0.10$^a$      &                     			&                          &                          &                    & 390 $\pm$ 70$^b$      \\
HD 204827   & B0V       & 1.11   $\pm$ 0.04$^a$      & 21.35 $\pm$ 0.05$^a$    &                          & 0.65    $\pm$ 0.10$^a$   &                    & 10  $\pm$ 10$^b$      \\
HD 208501   & B8Ib      & 0.75   $\pm$ 0.04$^a$      &                    			&    				   	   & 				&	  				& 97  $\pm$ 54$^b$      \\
BD +63 1964 & B0II      & 1.00   $\pm$ 0.03$^a$      & 21.70  $\pm$ 0.10$^a$   & 21.10 $\pm$ 0.18$^a$     & 0.33    $\pm$ 0.14$^a$   &     & 74  $\pm$ 3$^e$       \\
\hline
\hline
\multicolumn{8}{c}{} \\[-0.8em]
\multicolumn{8}{c}{Notes. $^a$\cite{Fan_2017}, $^b$\cite{Galazutdinov2021}, $^c$\cite{Cox_2014}, $^d$\cite{Jenniskens_1997} \& $^e$\cite{Cordiner_2017}.}
\end{tabular}
}
\end{table*}